\documentclass[sigconf]{acmart}

\AtBeginDocument{%
  \providecommand\BibTeX{{%
    \normalfont B\kern-0.5em{\scshape i\kern-0.25em b}\kern-0.8em\TeX}}}

\copyrightyear{2024}
\acmYear{2024}
\setcopyright{rightsretained}
\acmConference[MOBILEHCI Adjunct '24]{26th International Conference on Mobile Human-Computer Interaction}{September 30-October 3, 2024}{Melbourne, VIC, Australia}
\acmBooktitle{26th International Conference on Mobile Human-Computer Interaction (MOBILEHCI Adjunct '24), September 30-October 3, 2024, Melbourne, VIC, Australia}
\acmDOI{10.1145/3640471.3680450}
\acmISBN{979-8-4007-0506-9/24/09}

\usepackage{multirow}
\usepackage{graphicx}
\usepackage{subfigure} 
\usepackage{xcolor,soul}
\usepackage{array}
\usepackage{booktabs}
\usepackage{tabularx}

\begin{document}


\title[MultiSurf-GPT]{MultiSurf-GPT: Facilitating Context-Aware Reasoning with Large-Scale Language Models for Multimodal Surface Sensing}

\author{Yongquan Hu}
\orcid{0000-0003-1315-8969}
 \email{yongquan.hu@unsw.edu.au}
\affiliation{%
 \institution{University of New South Wales}
 \streetaddress{Kingsford}
 \city{Sydney}
 \state{NSW}
 \country{Australia}}

 \author{Black Sun}
\orcid{0009-0001-6708-8655}
 \email{blackthompson770@gmail.com}
\affiliation{%
 \institution{Southern University of Science and Technology}
 \city{Shenzhen}
 \country{China}}

  \author{Pengcheng An}
\orcid{0000-0002-7705-2031}
 \email{anpengcheng88@gmail.com}
\affiliation{%
 \institution{Southern University of Science and Technology}
 \city{Shenzhen}
 \country{China}}

 \author{Zhuying Li}
\orcid{0000-0001-5474-6949}
 \email{yongquan.hu@unsw.edu.au}
\affiliation{%
 \institution{University of New South Wales}
 \streetaddress{Kingsford}
 \city{Sydney}
 \state{NSW}
 \country{Australia}}

\author{Wen Hu}
\orcid{0000-0002-4076-1811}
 \email{wen.hu@unsw.edu.au}
\affiliation{%
 \institution{University of New South Wales}
 \streetaddress{Kingsford}
 \city{Sydney}
 \state{NSW}
 \country{Australia}}

\author{Aaron J. Quigley}
\orcid{0000-0002-5274-6889}
\email{aquigley@acm.org}
\affiliation{%
 \institution{CSIRO's Data61}
 \streetaddress{Kingsford}
 \city{Sydney}
 \state{NSW}
 \country{Australia}}

\renewcommand{\shortauthors}{Yongquan Hu et al.}

\begin{abstract}

Surface sensing is widely employed in health diagnostics, manufacturing and safety monitoring. Advances in mobile sensing affords this potential for context awareness in mobile computing, typically with a single sensing modality. Emerging multimodal large-scale language models offer new opportunities. We propose MultiSurf-GPT, which utilizes the advanced capabilities of GPT-4o to process and interpret diverse modalities (radar, microscope and multispectral data) uniformly based on prompting strategies (zero-shot and few-shot prompting). We preliminarily validated our framework by using MultiSurf-GPT to identify low-level information, and to infer high-level context-aware analytics, demonstrating the capability of augmenting context-aware insights. This framework shows promise as a tool to expedite the development of more complex context-aware applications in the future, providing a faster, more cost-effective, and integrated solution.

\end{abstract}



\maketitle

\section{Introduction}

Ubiquitous and mobile computing fundamentally transforms our interaction with the digital world, where context awareness plays a pivotal role in enriching user experience and improving operational efficiency \cite{weiser1991computer,abowd1999towards,baldauf2007survey}. Particularly, surface sensing — the capability to discern surface materials or categorize objects — is essential for understanding the immediate environment of a mobile device \cite{harrison2008lightweight,siewiorek2003sensay, zhang2018wall++}. This understanding aids devices in adjusting their functionalities based on context, whether optimizing screen readability based on ambient light conditions or human activity states \cite{yu2015sensing,hu2023exploring} or adjusting the interface when placed in different locations \cite{cheng2021semanticadapt}.

Despite advancements, a significant research gap exists in the unified processing of multimodal surface sensing data. Current methodologies often focus on specific types of data derived from different electromagnetic wave bands, such as visible light image \cite{hu2023microcam}, radar signal in the radio band \cite{yeo2016radarcat,elvitigala2023radarfoot} and multispectral data including infrared and ultraviolet light \cite{yeo2017specam,schrapel2021spectrophone}. Each method typically develops customized algorithms tailored for one modality without leveraging the potential synergies between them. This siloed approach limits the ability to process complex datasets where multiple modalities overlap, presenting a clear opportunity for a more integrated solution \cite{gellersen2002multi,salber2000context}.

Furthermore, traditional Artificial Intelligence (AI) methods, such as machine learning algorithms, for context awareness are typically confined to specific low-level tasks such as object recognition or material identification \cite{huang2022context,augusto2017survey}. However, these methods often lack the capability to perform high-level reasoning tasks that necessitate integrating various sensory inputs to offer comprehensive insights into a user's environment or behavior \cite{augusto2022contexts,hu2023exploringai}. Multimodal Large Language Models (LLMs) present new opportunities to amalgamate multi-dimensional surface sensory data, enhancing overall context awareness \cite{csepregi2021effect}. LLMs excel in synthesizing information from diverse sources and generating rich context insights, predominantly based on processing and assimilating extensive generalized world knowledge \cite{csepregi2021effect}. It is crucial to note that despite the propensity of LLMs to generate insights with potential biases and inaccuracies, their capacity for logical and causal reasoning remains impressive \cite{yuan2024r,coletta2024llm}. As technology evolves, these inherent risks are expected to diminish, which falls outside the purview of this study.

To address these challenges, we introduce MultiSurf-GPT, a framework leveraging the latest GPT-4o model to process multimodal surface sensing data (radar, microscope and multispectral data), aiming to establish a unified workflow from recognition to reasoning for high-level context awareness. Notably, the current capabilities of LLMs are primarily tailored to text, image, and file modalities \footnote{GPT-4o: https://openai.com/index/hello-gpt-4o/}. Currently, how to uniformly process multimodal sensor data using LLMs remains an open question. Optimizing LLMs’ recognition of sensor data typically requires instruction tuning, which is resource-intensive \cite{ouyang2024llmsense}. Alternatively, converting all data to the native modalities supported by GPT-40, such as images or CSV files, and using the LLM as an interface to invoke existing classification algorithms based on prompt engineering provides a cost-effective and efficient solution for rapid prototyping \cite{mirza2024meta,berenguer2024leveraging}. In this study, we adopt the prompt-based strategies (zero-shot prompting and  prompting) for rapid prototyping, but we anticipate incorporating instruction fine-tuning in future work.

The main contributions of this paper are threefold: (1) We introduce MultiSurf-GPT, using GPT-4o as an agent within a unified 2R framework for enhanced contextual information reasoning. We consider it as a tool to advance the future development of more adaptive and intelligent mobile applications; (2) We demonstrate the application of LLMs in the unified processing of multimodal surface sensing data, with our test results establishing a benchmark for further work, including case studies on high-level reasoning from diverse contexts; (3) We briefly dicuss the limitations and future directions, providing insights for ongoing efforts. Overall, this research not only broadens the technological landscape but also opens new avenues for the practical application of LLMs in ubiquitous computing.

\section{Related Work} \label{related work}

\subsection{Multimodal Surface Sensing Data for Ccontext Awareness}
The analysis of multimodal surface sensing data is key to context awareness. With advancements in AI, surface sensing has become miniaturized and mobile. Based on electromagnetic band distribution, surface sensing data can be categorized into image (visible light band), multispectral (usually a combination of multiple invisible light bands such as infrared and ultraviolet), and radar (radio band). Firstly, visible images, such as the grayscale images used by Yang et al. in MagicFinger \cite{yang2012magic} or the RGB images used by Hu et al. in MicroCam \cite{hu2023microcam}, offer high resolution for detailed observation at a micro or nano scale, enhancing interpretability and expanding interactive possibilities. However, they suffer from a limited field of view and require extensive sample preparation. Secondly, multispectral data excels in differentiating materials based on spectral signatures, but it requires complex analysis and high equipment costs, making it uncommon on mobile devices. For example, Harrison et al. used wearable multispectral hardware for material classification \cite{harrison2008lightweight}, and Yeo et al. achieved high recognition accuracy with this sensor on a mobile device in SpeCam \cite{yeo2017specam}. Lastly, radar data can penetrate obstacles and cover large areas, useful in wide applications such as meteorology and surveillance, but usually offers lower resolution compared to optical methods and requires a clear line-of-sight. For instance, Yeo et al. used Google's Soli sensing suite in RadarCat for surface material classification \cite{yeo2016radarcat}. Yeo et al. further extended radar-based context awareness to the recognition of tangible surfaces in Tangible Radar \cite{yeo2018exploring}. Additionally, Samitha et al. employed higher resolution radar signals in RadarFoot for fine-grained ground recognition and classification. \cite{elvitigala2023radarfoot}. In summary, current methods for context-aware surface sensing primarily concentrate on processing data from a single modality, with limited efforts directed towards the comprehensive understanding of multimodal data to deliver intricate contextual information about the environment.

\subsection{LLM Applications in Sensing Data Processing and Reasoning}
Recent advances in multimodal LLMs have significantly enhanced their capability to process sensor data, catalyzing a transformative shift across various domains, including environmental monitoring and autonomous systems. A prime example is EarthGPT, which integrates multimodal data streams such as optical, Synthetic Aperture Radar (SAR), and infrared data to improve the understanding of remote sensing images \cite{zhang2024earthgpt}. Similarly, VLMRemote utilizes multimodal LLMs to merge visual and linguistic modalities in remote sensing, enhancing geographic feature recognition by combining textual descriptions with visual data \cite{li2024vision}. Moreover, multimodal LLMs demonstrate robust reasoning abilities over low-level sensor data. The LLMSense framework, for example, advances high-level reasoning over spatiotemporal sensor trajectories by leveraging multimodal LLMs' inherent capacity to assimilate extensive world knowledge and reasoning capabilities, effectively interpreting long-term environmental and situational dynamics \cite{ouyang2024llmsense}. HealthLLM employs multimodal LLMs to analyze health-related sensor data, diagnosing and predicting health conditions by integrating physiological data with contextual information for actionable healthcare insights \cite{kim2024health}. In short, these developments highlight the potential of multimodal LLMs in the unified processing of multimodal sensor data. Their ability to interpret and reason about high-level information from low-level data is crucial for managing complex sensor inputs and paves the way for the development of smarter, more adaptive technologies.

\section{Methodology}\label{methodology}

\subsection{Dataset Selection}

For context-aware surface sensing in mobile computing, we selected three datasets of different modalities: (1) Tangible Radar \cite{yeo2018exploring} dataset (radar data as CSV file): This dataset consists of radar signal recordings used to detect and classify physical objects within an environment. This dataset comprises radar reflection signals collected by a single-mode, multi-channel (8 channels, 2 transmitters, and 4 receivers) system. These signals encompass information about an object's distance, thickness, shape, density, and internal composition. (2) MicroCam \cite{hu2023microcam} dataset (microscope image): This dataset comprises high-resolution images captured using microscopic cameras with the phone face-up, specifically designed to detail the textural and material properties of various surfaces. It is a valuable resource for researchers focusing on fine-grained image analysis and material identification. (3) SpeCam \cite{yeo2017specam} dataset (multispectral image): The data contains reflection light images, collected by placing a smartphone screen facing down and rapidly flashing different colors on the screen while the front camera captures the reflected light based on spectral cameras. This reflected light contains information about the surface material's color and optical properties.

\subsection{Prompt Design}

For our methods in MultiSurf-GPT framework, we utilize prompt engineering to address multiple tasks across various datasets as an initial exploration. These prompts are model-agnostic, and detailed descriptions of the language models and experimental settings will be provided in the following section. Our prompts are crafted to handle different modalities and ensure flexibility. Additionally, we have assessed and compared the effectiveness of zero-shot and  prompts in our evaluations.

It is worth emphasizing that the processing of MultiSurf-GPT is not confined to specific datasets. After constructing a model to analyze a dataset, MultiSurf-GPT also review the description document (published papers) associated with it, extracting critical information from these documents, such as the sensing methods and the usage domains. Related details could be found in the ``Extra settings of all datasets'' part of the Section \ref{experiment}.

\noindent \textbf{Zero-shot prompting.} As shown in Table~\ref{tab: prompt design}, the zero-shot prompting strategy consists of a specific task explaining, an additional rule to avoid unnecessary output, and restricted models to focus on the current task. Therefore, the final prompt for the model consisted of \{Task Explaining\} + \{Rules\}.

    \noindent\textbf{1-shot prompting.}The  prompt added the  samples after the same zero-shot prompt template (We only use a one-shot prompt). Specifically, we include the task explaining prompt following the zero-shot prompt but provide the correct class labels instead of offering different candidate class labels for prediction.

\begin{table*}[]
\centering
\caption{The zero-shot and  prompting strategies. <MOD> as a placeholder denotes different modalities (such as radar). <MODEL> as a placeholder denotes different Machine Learning Models (such as SVM, RF). <CLASS> as a placeholder denotes the list of categories of pictures. <PIC1>, <PIC2>...<PICn> as placeholders denote the example for each category. As for zero-shot, the <PIC1>, <PIC2>...<PICn> examples will be deleted.
}
\label{tab: prompt design}
\begin{tabular}{@{}l>{\raggedright\arraybackslash}p{0.6\linewidth}>{\raggedright\arraybackslash}p{0.25\linewidth}@{}}
\toprule
Data    & Task Explaining & Rules                     \\ \midrule
CSV file     & The provided CSV is <MOD> data. In the CSV file, columns [0:-1] contain the radar features, and the last column [-1] contains the labels. Build a model (defaulting to using <MODEL>) and return the accuracy. Do not output other text.& [Rules]: Do not output any other text.\\ \midrule
Image & The provided picture is <MOD>. Identify the category of this picture from <CLASS> and return only one category (only one category can be returned). For example, <PIC1>, <PIC2>...<PICn> (where n is the number of categories in <CLASS>) are sample images for each category in <CLASS>.& [Rules]: Must return within <CLASS>. Do not output any other text.\\ \midrule
Document (paper)   & According to the given paper, what equipment did they use, what is the method, and what is the origin usage of the data? & [Rules]: Summarize the method and origin usage each in one complete sentence.\\ \bottomrule
\end{tabular}
\end{table*}

\section{Experiment}\label{experiment}

\begin{figure*}[htbp]
    \centering
    \includegraphics[width=\textwidth]{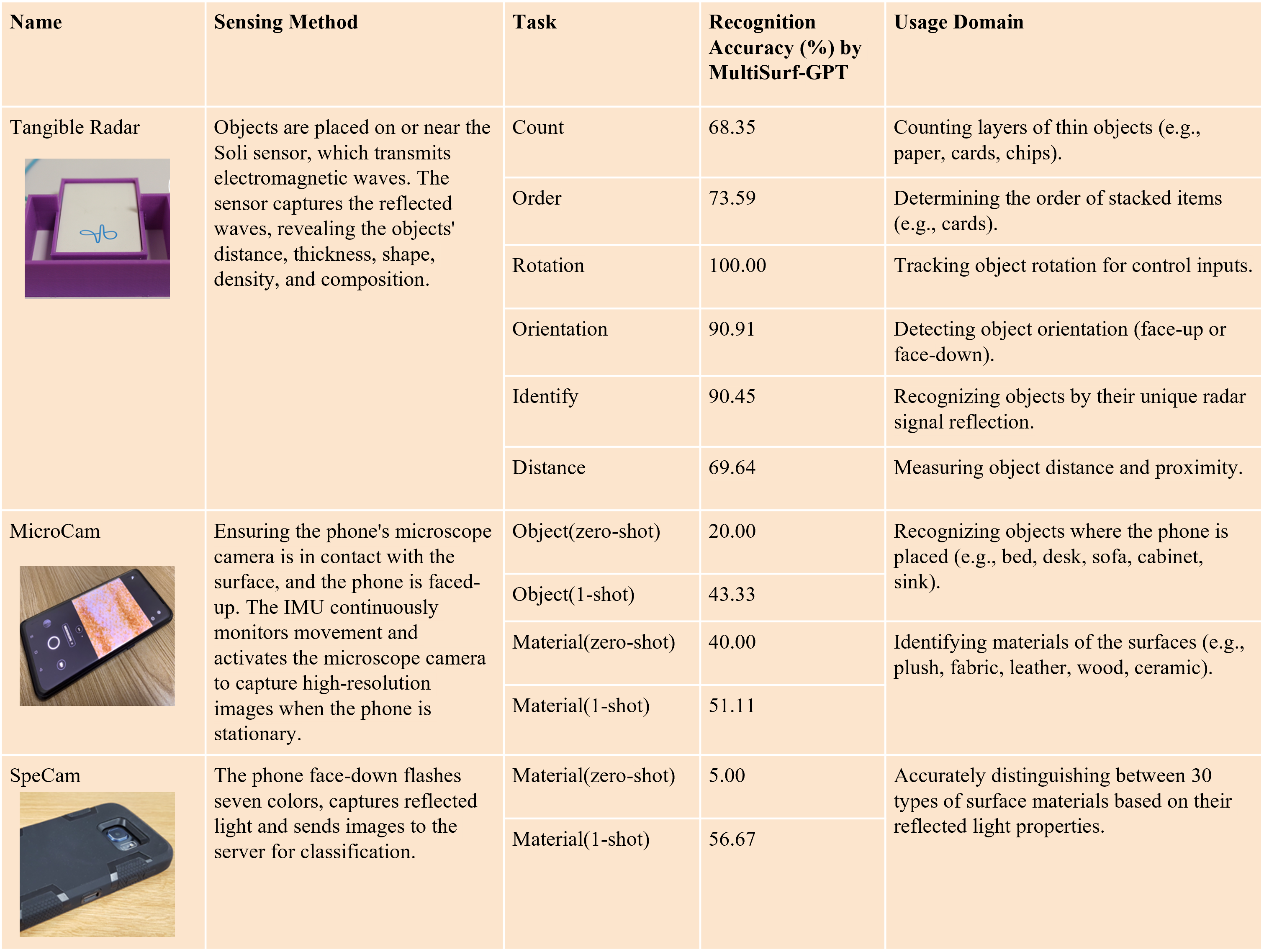}
    \caption{Low-level information on 3 surface sensing methods as recognized and captured by the MultiSurf-GPT framework for high-level context-aware reasoning.}
    \label{fig:result1}
\end{figure*}

\subsection{Settings}

\subsubsection{Dataset Settings}


We leverage several input settings for different modalities data: (1) Tangible Radar dataset settings: we input a CSV file containing radar signals and class labels. This utilizes GPT-4o's ability to read CSV files (file modality) and employ the code interpreter to analyze the data using Python code for methods such as SVM \cite{hearst1998support}. (2) MircoCam dataset settings \cite{hu2023microcam}: we input microscope images (image modality), utilizing GPT-4o's image modality recognition capabilities. For  prompts, we provide one example image per category for GPT-4o to recognize. In contrast, no examples are given for zero-shot prompts. (3) SpeCam dataset settings \cite{yeo2017specam}: we input multispectral images (image modality), again leveraging GPT-4o's image modality recognition abilities. Like the MircoCam settings, we provide one example image per category for  prompts and none for zero-shot prompts. (4) Extra Settings of all datasets: the original research papers of the three datasets are also inputted as files, utilizing GPT-4o's capabilities to retrieve, read, and comprehend the documents.

\subsubsection{Model Settings}

Given that GPT-4o is the most recent series of GPT-4 that naturally supports multimodal capabilities, we adopted GPT-4o as the tested LLMs.  For the  experiment, we tested the  learning scenario to examine the capability of multi-model LLMs with limited information provided. We randomly selected one sample from the corresponding dataset as the  sample in each repeated trial.  We use the image updating module of GPT-4o. However, we use no other additional techniques (e.g., Chain-of-Thoughts \cite{wei2022chain}) to serve as a preliminary study of how multimodal LLMs process multimodal information. This approach ensures the results reflect the basic capability of the models, which was also consistent with previous work \cite{xu2023leveraging}.

\begin{figure*}[htbp]
    \centering
    \includegraphics[width=0.95\textwidth]{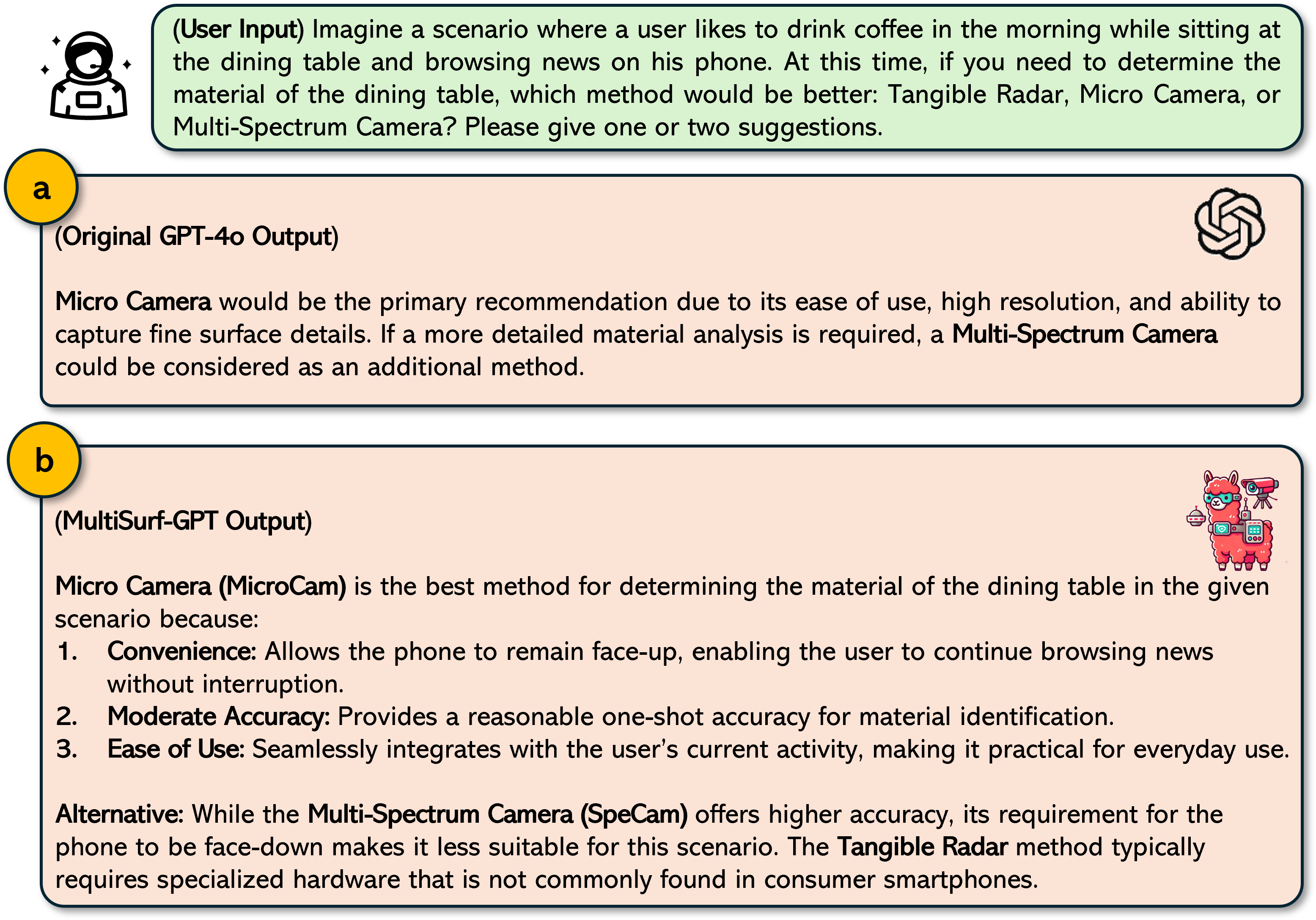}
    \caption[Comparison of Method Recommendations for Surface Material Identification by Different Models]{A good experiment case: (a) represents the output by the original GPT-4o model, recommending a Micro Camera as the primary method and a Multi-Spectrum Camera for more detailed analysis if needed. (b) illustrates the output by MultiSurf-GPT, which favors the Micro Camera for its convenience, moderate accuracy, and ease of use, specifically highlighting the practical limitations of the Multi-Spectrum Camera and Tangible Radar for everyday scenarios.}
    \label{fig:compare}
\end{figure*}

\subsection{Results and Discussions}

\subsubsection{Performance of MultiSurf-GPT}

Regarding the Tangible Radar dataset, MultiSurf-GPT demonstrated remarkable proficiency in generating machine learning code to analyze CSV data. Specifically, the tasks of Rotation, Orientation, and Identification achieved impressive accuracy rates of 100\%, 90.91\%, and 90.45\%, respectively. However, Count, Order, and Distance tasks exhibited comparatively lower performance. This discrepancy is attributed to the inferior performance of the Random Forest (RF) model used by the code interpreter rather than the inherent capabilities of GPT-4o itself. For the MicroCam and SpeCam datasets, GPT-4o leveraged its image recognition capabilities. In zero-shot scenarios, GPT-4o struggled to identify images due to their limited semantic information accurately. Nonetheless, because of the high similarity between images, the accuracy of GPT-4o's recognition improved significantly in one-shot scenarios. Specifically, for the MicroCam dataset, the accuracy in Object and Material tasks increased by 23.33\% and 11.11\%, respectively. For the SpeCam dataset, the accuracy rose impressively by 51.67\%. Moreover, MultiSurf-GPT excels in extracting detailed information from original research papers, such as the sensing methods and usage domains. These capabilities establish a robust foundation for more complex, context-aware analyses in future applications.

\subsubsection{Case Analysis of Context-Awareness}

As shown in Figure \ref{fig:compare}, the output is generated separately by GPT-4o and MultiSurf-GPT. As Dey et al. mentioned \cite{abowd1999towards}, context awareness encompasses four aspects: location, activity, time, and identity. The original GPT-4o provided overly simplistic answers and failed to address these four aspects adequately. In contrast, MultiSurf-GPT delivers more interpretable results by utilizing user input, datasets, and the context-awareness of the papers. For instance, MultiSurf-GPT correctly interprets that ``browsing news on his phone'' requires the phone to be face-up, aligning with MicroCam's method of using a micro camera on the phone. Although the Multi-Spectrum Camera (SpeCam) offers higher accuracy, its requirement for the phone to be face-down makes it less suitable for this scenario. MultiSurf-GPT seamlessly integrates with the user's current activity, making it practical for everyday use. This demonstrates a thorough understanding of the ``activity'' context, consistent with Byun et al.'s research \cite{byun2001exploiting}. Regarding the statement, ``The Tangible Radar method typically requires specialized hardware not commonly found in consumer smartphones,'' MultiSurf-GPT recognizes that consumer smartphones generally lack radar devices, indicating a solid grasp of the ``identity'' context. Therefore, MultiSurf-GPT performs better than the original GPT-4o in terms of context awareness.

\section{Conclusion and Future Work}

We have investigated the integration of multimodal LLMs like GPT-4o through our MultiSurf-GPT framework for processing and interpreting multimodal surface sensing data across radar, microscopy, and multispectral modalities. Utilizing zero-shot and one-shot cueing strategies, MultiSurf-GPT has shown initial success in recognizing low-level information and deriving high-level contextual insights. This deployment of LLMs in mobile contextual awareness paves the way for more sophisticated and integrated applications, offering enhanced user-device interactions through advanced reasoning and data processing capabilities.

In the future, the MultiSurf-GPT framework presents several opportunities for enhancement and further research. Currently, its reliance on cue engineering can limit recognition accuracy, particularly with complex datasets like MicroCam and SpeCam. Future versions could benefit from incorporating instruction fine-tuning to improve recognition accuracy. Additionally, wide-range subjective user experiments would help quantitatively and precisely assess how MultiSurf-GPT contributes to enhanced contextual awareness in surface sensing. These initiatives would not only refine the model's performance but also broaden its practical applications, fostering the development of more adaptive and intelligent mobile applications.


\bibliographystyle{ACM-Reference-Format}
\bibliography{sample}


\end{document}